\documentclass[a4paper]{article}
\pdfoutput=1
\usepackage{xcolor}
\usepackage{tabularx}
\usepackage{bbm}
\usepackage[colorlinks]{hyperref}
\usepackage[utf8]{inputenc}
\usepackage[T1]{fontenc}
\usepackage{float}
\usepackage{lipsum}
\usepackage{subcaption}

\definecolor{britishracinggreen}{rgb}{60,179,113}
\definecolor{darkgreen}{rgb}{0.0, 0.2, 0.13}
\usepackage{INTERSPEECH2019}

\title{
Vocal markers from sustained phonation in Huntington's Disease
}
\name{Rachid Riad$^{1,2}$, Hadrien Titeux$^1$, Laurie Lemoine$^{2,3}$,  Justine Montillot$^{2,3}$, Jennifer Hamet Bagnou$^{2,3}$, Xuan Nga Cao$^1$, Emmanuel Dupoux$^{1}$, Anne-Catherine Bachoud-Lévi$^{2,3}$ }
\address{
  $^1$ CoML/ENS/CNRS/EHESS/INRIA/PSL Research University, Paris, France \\
  $^2$ NPI/ENS/INSERM/UPEC/PSL Research University, Paris, France  \\
  $^3$ Huntington's Disease National Reference Center, Neurology Department, \\
   Henri-Mondor Hospital, APHP, Cr\'eteil, France
  }
\email{rachid.riad@ens.fr}

\begin{document}

\maketitle

\begin{abstract}


Disease-modifying treatments are currently assessed in neurodegenerative diseases. Huntington's Disease  represents a unique opportunity to design automatic sub-clinical markers, even in premanifest gene carriers. We investigated phonatory impairments as potential clinical markers and propose them for both diagnosis and gene carriers follow-up. We used two sets of features: Phonatory features and Modulation Power Spectrum Features. We found that phonation is not sufficient for the identification of sub-clinical disorders of premanifest gene carriers. According to our regression results, Phonatory features are suitable for the predictions of clinical performance in Huntington's Disease.

\end{abstract}
\section{Introduction}

Huntington’s disease is an autosomal dominant neurodegenerative disease with a complete penetrance \cite{gusella1983polymorphic}. The disease is characterised by a triad of symptoms (motor, cognitive and psychiatric \cite{novak2010huntington}) that leads to the progressive disability functioning of the individual with HD. The age of the clinical onset is highly variable (mean around 45 year-old) and the symptoms gradually worsen over 15 to 20 years until the death.
The motor assessment allows defining a cut-off score that splits between premanifest and manifest gene carriers and thus the disease onset. Therefore, some of the deficits cognitive or psychiatric impairments or sub-clinical disorders may appear long before the disease manifests.

The automatic identification of pre-symptomatic carriers of the gene of  Huntington's Disease and the prediction of the clinical scores are of particular interest for the neurological practice \cite{wild2019one}. Indeed, having ecological markers of these clinical endpoints may: (1) help to prevent detrimental and harmful life events, (2) speed-up clinical trials, (3) increase the understanding of the disease in ground truth condition.
Here, by using speech analysis of sustained phonation, we aim at distinguishing, in individuals, stage of the disease and predicting their clinical scores, as assessed by neurologists and neuropsychologists.

We rely on the fact that individuals with Huntington's Disease can exhibit different disorders during speech \cite{rusz2014phonatory, vogel2012speech} and language production \cite{hinzen2018systematic, teichmann2005role}.
Motor speech disorders in Huntington's Disease are commonly referred to as \textit{hyperkinetic} dysarthria \cite{darley1969differential}: variable rate, abnormal prosody, imprecise consonants and distorted vowels, phonation deviations, and sudden forced breath. Yet, there is some recent evidence that the speech motor impairments in Huntington's Disease are highly heterogeneous \cite{diehl2019motor}.

Here, we considered the speech features collected from simple recordings of the French vowel /a/ uttered in a sustained manner for as long as possible, in regular clinical conditions. We collected production of these vowels for Healthy Controls (C), gene carriers without overt manifestation of Huntington's Disease (preHD) and manifest gene carriers of Huntington's Disease (HD). We modelled each vowel with several features, which are impaired in hyperkinetic dysarthria. We investigated two sets of features: (1) Phonatory features already proven useful to distinguish gene carriers from control (preHD vs Controls \cite{rusz2014phonatory} and HD vs Controls \cite{rusz2013objective}), (2) Modulation Power Spectrum features to measure the modulations that characterise speech intelligibility \cite{elliott2009modulation} and roughness \cite{arnal2015human}. First, we conducted a statistical analysis at the group level of these features. Then, based on these sets of features, we used regularised linear models to predict the group and clinical scores of the patients.

Our statistical analyses and classification results showed that HD patients distinguished from controls, whereas the boundaries around preHD are more blurred. The Modulation Power Spectrum features complemented the Phonatory features to help identify preHD and improve the F1-score.
In contrast, we observed that the Phonatory features have the best predictive capabilities of the clinical scores.

\section{Related work}
Previous studies have used sustained phonation for the assessment of various neurodegenerative disorders such as Parkinson's disease \cite{little2007exploiting}, Amyotrophic Lateral Sclerosis  \cite{peplinski2019objective} and also Huntington's Disease \cite{ rusz2013objective}. These studies have extracted a number of hand-crafted features from the sustained phonation and built discriminative models to distinguish patients from controls. Yet, collecting pre-symptomatic speech data on other acquired neurodegenerative diseases is difficult, and can only be done retrospectively \cite{berisha2017float}.

\begin{table}[b]
\vspace{-1em}
\caption{Participants demographics and clinical scores}
	\centering

	 \begin{tabular}{|p{0.23\linewidth}|c|c|c|}

		\hline
		 & Controls & \multicolumn{2}{c|}{Huntington's disease } \\
		 &  & \multicolumn{2}{c|}{Gene carriers} \\
		 \hline
		                Sub-groups            & C  & PreHD     &  HD          \\
		\hline

		N               & 24  & 16 & 45       \\

		         Gender        & 12F/12M   & 9F/7M  & 27F/18M    \\
				Age  (years)                 &  54.1 (8.9)  & 50.2 (12.2) & 53.9 (11.3)      \\
				 CAG Triplets                   & $\leq$ 35  & 41.4 (1.5) & 44.3 (3.4)       \\
				 \hline
				 cUHDRS \cite{schobel2017motor}                  & ---  & 17.6 (1.4) & 9.0 (3.6)       \\
				 TFC     \cite{shoulson1981huntington}             & ---  & 13.0 (0.0) & 10.3 (2.2)       \\
					TMS               & ---  & 0.38 (1.0) & 36.0 (15.9)       \\
		\hline
	\end{tabular}
	    \label{tab:demographic_patients}
\vspace{-1em}
\end{table}

Rusz et al. \cite{rusz2014phonatory} provided the closest study to ours using sustained phonation for the automatic identification of preHD. They show great discriminative performance between preHD and controls.

Perez et al. \cite{perez2018classification} trained a Speech Recognition system on Huntington's speech and extracted a number of language features ranging from speech rate, and goodness of pronunciation to utterance length. The goal of this study differed from ours as they grouped together the preHD and HD patients in a single group to distinguish from control. They also pointed out to the difficulty to identify preHD.

We did not find any study attempting to predict the clinical scores in Huntington's disease from speech. Our strategy, which may allow following up individuals remotely, is likely to apply to other neurodegenerative diseases, such as in the Parkinson's Disease \cite{smith2017vocal} or Alzheimer Disease \cite{LuzHaiderEtAl20ADReSS}.

\section{Voice Database}
Eighty five participants were included from two observational cohorts (NCT01412125 and NCT03119246) in this ancillary study at the Hospital Henri-Mondor Créteil, France): 61 people tested  with a number of CAG repeats on the Huntingtin gene above 35 \cite{gusella1983polymorphic} (CAG $>35$), and 24 Healthy Controls (C) (See Table \ref{tab:demographic_patients}). All participants signed an informed consent.
Mutant Huntingtin gene carriers were considered premanifest if both they score less than five at the Total Motor score (TMS) and their Total functional capacity (TFC) equals 13 \cite{tabrizi2009biological} using the Unified Huntington Disease Rating Scale (UHDRS) \cite{kieburtz2001unified}.

All participants completed a standardised speech battery. The data were annotated with Seshat \cite{seshat2020titeuxriad} and Praat \cite{boersma2002praat} softwares. The annotators were second-year graduate students in speech pathology, all French native speakers.

Each participant was asked to take a deep breath and to sustain the vowel /a/ at a constant intensity and pitch level for as long as possible. The recordings were done in the same condition for all participants, with a ZOOM H4n Pro recorder, sampled at 44.1 kHz with a 16-bit resolution.

\begin{table}[!ht]

	\caption{List of Phonatory features based on \cite{rusz2014phonatory}. SD stands for Standard Deviation.
	$\dagger$\footnotesize{The vocal tremor features could not be computed on all samples. }}
	\centering
	\begin{tabular}{|l|>{\footnotesize}c|}
		\hline
		Dimension  & Features  \\
		\hline
		Airflow insufficiency   & Maximum Phonation Time               \\
		         & First Occurrence of Voice Break             \\
		\hline
		Aperiodicity   & Number of Voice Breaks            \\
		         & Degree of Pitch Breaks           \\
		         		         & Degree of Vocal Arrests
          \\
		\hline
		Irregular vibration    & F0 SD            \\
		   of vocal folds      & Recurrence Period Density Entropy          \\

		\hline
		Signal perturbation   & Jitter (local)    \\
		         & Shimmer     (local)     \\

		\hline
		Increased noise   & Harmonics to Noise Ratio     \\
		         & Detrended Fluctuation Analysis
     \\

		\hline
		Vocal tremor$\dagger$   & Frequency Tremor Intensity Index     \\
		         & Amplitude Tremor Intensity Index
     \\

		\hline
		Articulatory deficiency
   & Mean of SD of MFCC     \\
		         & Mean of SD of Delta MFCC
     \\

		\hline
	\end{tabular}
 	    \label{tab:vocal_features}

\end{table}

\vspace{-1.5em}

\section{Features}
\subsection{Phonatory Features}

We used the Phonatory features from \cite{rusz2013objective,rusz2014phonatory} to measure dimensions that can impede the correct sustained production of the vowel /a/ for HD and preHD gene carriers: airflow insufficiency,  aperiodicity, irregular vibration of vocal folds, signal perturbation, increased noise, vocal tremor, articulatory deficiency. These features are summarised in the Table \ref{tab:vocal_features}.

To automatically extract these features, we used the Praat software \cite{boersma2002praat}, the Parselmouth wrapper \cite{parselmouth} and the tremor package \cite{bruckl2012vocal}. The fundamental frequency (F0) and MFCCs were obtained with the Kaldi toolkit \cite{ghahremani2014pitch}. Besides, we implemented the Detrended Fluctuation Analysis and the Recurrence Period Density Entropy features from \cite{little2007exploiting}, in Python. To compute these two features and replicate \cite{little2007exploiting}, we down-sampled the audio to 22.5 kHz, otherwise the audio is down-sampled to 16 kHz.
Due to instability for the extraction of the tremor features, there are missing data points for this dimension (See Table \ref{tab:results_stats}).

To overcome the limitations of heterogeneity of acoustic methodologies across studies and libraries, we provide an open-source version of the code (\href{https://github.com/bootphon/sustained-phonation-features}{link}) to reproduce our results and to extract each dimension of the sustained phonation of the vowel /a/.

\subsection{Modulation Power Spectrum features}

To complete the Phonatory features, we used the Modulation Power Spectrum (MPS) extracted for each vowel /a/ of each individual. Different perceptual attributes occupy distinct areas of the MPS: roughness, gender and size characteristics of the speaker, and linguistic meaning \cite{arnal2015human}. Besides, the MPS captures the spectral and amplitude modulations of the pitch and its harmonics \footnote{Readers can refer to \cite{elliott2009modulation,arnal2015human} for in-depth study of the different perceptual areas}.
The MPS representation is the amplitude spectrum of the 2D Fourier Transform of a time-frequency representation obtained from the sound waveform \cite{elliott2009modulation,elie2018zebra}. This time-frequency representation is a log-scaled amplitude of a spectrogram computed every 1 millisecond (ms)  with a spacing between each frequency bin of 50Hz using a Gaussian window. The linear spacing in the frequency axis \cite{elliott2009modulation} better describes sounds that have harmonic structure, like long steady vowels. The MPS is computed every 10 ms with a 100 ms window of the spectrogram, then the final representation is averaged over the full duration of the sound. Upward and downward temporal modulations are kept between -200 Hz and +200 Hz, and spectral modulations between 0 and 9.5 Cycles/kHz (99\% of the energy was found between these intervals).

\section{Methods}

\begin{table*}
	\centering
		\caption{Results of the statistical analyses between the three groups for the sustain of the vowel /a/: Controls (C), asymptomatic genetic carrier of Huntington's Disease (preHD), symptomatic genetic carrier of Huntington's Disease (HD). The p-values significativity of the tests (Kruskal-Wallis tests H-statistic (H stat) and post-hoc pairwise tests on the Cohen's d) are reported with *: $0.01 < p \leq 0.05$, **: $0.001 < p \leq 0.01$, ***: $ p \leq 0.001$. The post-hoc statistics are corrected for multiple comparison feature wise. SD stands for standard deviation.
		$\dagger$ 23 HD, 6 preHD, 6 Controls data points could not be computed for the Frequency Tremor Intensity Index. $\ddagger$ 15 HD, 6 preHD, 3 Controls data points could not be computed for the Amplitude Tremor Intensity Index.
}
	\begin{tabular}{>{\footnotesize}p{0.24\linewidth}|p{0.083\linewidth}p{0.083\linewidth}p{0.083\linewidth}|p{0.06\linewidth}|p{0.07\linewidth}p{0.07\linewidth}p{0.07\linewidth}}
		\hline
		  & \multicolumn{3}{c|}{Mean (SD)} & \multicolumn{1}{c|}{H stat}  & \multicolumn{3}{c}{Effect size Cohen’s d}\\
		\hline

		                          & C   & preHD     & HD         &  & HD/PreHD     & HD/C    & preHD/C    \\
		\hline
		\hline
		\multicolumn{8}{l}{\textit{Phonatory Features}}  \\
		\hline
		Maximum Phonation Time      (s)      & 17.0 (8.7)     & 15.7 (4.9) & 9.1 (5.1)  & 23.6$^{***}$      & -1.27$^{***}$  & -1.19$^{***}$ & -0.18      \\

		First Occurrence of Voice Break   (s)                    & 13.2 (7.6) & 10.3 (5.7) & 6.3 (5.1)  &16.6$^{***}$      & -0.77$^{**}$ & -1.14$^{***}$ & 	-0.41     \\
\hline
		Number of Voice Breaks             & 3.6 (9.2)        & 7.2 (12.3) & 3.8 (7.4) & 2.9     & -0.37 & 0.02 & 	0.34     \\

		Degree of Pitch Breaks  (\%)    & 1.1 (3.1)      & 4.8 (9.4)& 6.3 (10.6)  & 5.7     & 0.15 & 0.59$^{**}$ & 0.59       \\
		Degree of Vocal Arrests  (\%)  & 1.6 (2.8)  & 3.8 (5.9) &7.3 (10.1) & 6.8      & 0.39 & 0.68$^{**}$  & 0.49     \\
		\hline
		F0 SD    (Hz)   & 7.3 (10.8)   & 	13.8 (15.2) & 16.8 (15.3)  & 13.1$^{**}$      & 0.20 & 0.68$^{***}$ & 0.51      \\

		Recurrence Period Density Entropy                      & 0.60 (0.18)  & 0.57 (0.16) & 0.56 (0.16) & 0.52     & -0.09 & -0.23 & -0.13     \\
		\hline
		Jitter (local)        (\%)       & 0.73 (0.47)      & 1.1 (1.7) & 1.0 (1.0) & 0.70     & -0.10 & 0.32 & 0.36    \\

		Shimmer (local)      (\%)   & 8.5 (3.6)    & 7.9 (4.0)& 8.4 (4.5) & 0.37      & 0.14 & 0.02 & 0.15       \\
		\hline
		Harmonics to Noise Ratio  (dB) & 16.1 (4.1)  & 15.8 (5.2) & 15.3 (5.1)  & 0.52      & -0.11 & -0.19 &  	-0.07      \\
		Detrended Fluctuation Analysis    & 0.70 (0.04)     & 0.70 (0.04)& 0.68 (0.06)  & 2.5      & -0.52 & -0.41 & 0.16      \\
		\hline
		Frequency Tremor Intensity Index $\dagger$ & 4.0 (4.3) & 7.1 (6.4) & 7.4 (8.1)  & 3.5      & 0.07 & 0.52 & 0.60      \\
		Amplitude Tremor Intensity Index  $\ddagger$ & 24.2 (7.6) & 29.4 (14.7) & 27.2 (14.4)  & 13.7$^{**}$     & 0.57 &1.10 & 0.50      \\
		\hline
		Mean of SD of MFCC & 6.6 (1.2) & 6.6 (0.79) & 8.5 (1.7)  & 25.6$^{***}$      & 1.23$^{***}$  &1.21$^{***}$ & 0.01     \\
		Mean of SD of Delta MFCC & 1.3 (0.19) & 1.2 (0.25) & 1.5 (0.38)    & 14.7$^{**}$     & 0.80$^{**}$ & 	0.77$^{***}$  & -0.13      \\
		\hline
		\hline

	\end{tabular}

	    \label{tab:results_stats}
\vspace{-1em}
\end{table*}

\subsection{Group level: Statistical analysis}


We followed the process for statistical comparison and correction from Rusz et al., 2014 \cite{rusz2014phonatory}. Given the non-normality of the data, we tested the differences between groups with the non-parametric Kruskal-Wallis test. The Bonferroni correction is applied to correct for the seven types of speech deficits (see Table \ref{tab:vocal_features}).
For post-hoc analyses, we applied the Kolmogorov-Smirnov test to all features to check for normality within each group, and the Levene's test for homoscedasticity between groups. If normality and homoscedasticity requirements were fulfilled, we applied an independent t-test otherwise; we applied a non-parametric equivalent (Mann-Whitney U-test). The p-values were Bonferroni corrected for multiple comparisons per feature.
We also estimated the effect size with the Cohen's d. Results are summarised in Table \ref{tab:results_stats}. Statistical analyses for the MPS features are displayed in the arxiv version of this paper.

\subsection{Individual level: Machine Learning}
To assess the performance of each set of features for classification and regression, we conducted 100 repeated learning-testing with 20\% of the data left out as test set \cite{varoquaux2017assessing}. We used scikit-learn for all our models and data processing \cite{scikit-learn}.
\subsubsection{Group classification}

 We first compared the predictive power of the different input features to discriminate between the three groups: Controls (C), preHD, and HD patients. To do so, we trained a logistic regression regularised with ElasticNet (e.g. L1 and L2 combined with $C=1$ and $ratio=0.5$). Then, we computed the mean and standard deviation of the Accuracy and the F1 score (See Table \ref{tab:results_ML}). We also reported the chance level by selecting randomly the class based on the prior distribution.
 \subsubsection{Regression of the clinical scores}

 Second, we assessed the predictive capabilities of the features to predict the composite score cUHDRS \cite{schobel2017motor} (currently used in international clinical trials \cite{tabrizi2019targeting}), the TFC and the TMS. We trained a Linear regression model a with ElasticNet regulariser (e.g. L1 and L2 combined with $C=1$ and $ratio=0.5$). The phonation from Controls are not used in the regression. Then, we computed the mean and standard deviation of the Mean Absolute Error and the coefficient of Determination $R^2$ (See Table \ref{tab:results_ML}) as well as the chance level by predicting the mean of each outcome on the train set.

\section{Results and discussions}






\begin{figure*}[!ht]
    \centering
    \caption{Averaged weights of the Logistic Regression regularised with ElasticNet applied on the Modulation Power Spectrum Features to discriminate between each sub-group. Mean Sparsity $ = 37.1\% $
    }
    \includegraphics[width=0.9\textwidth]{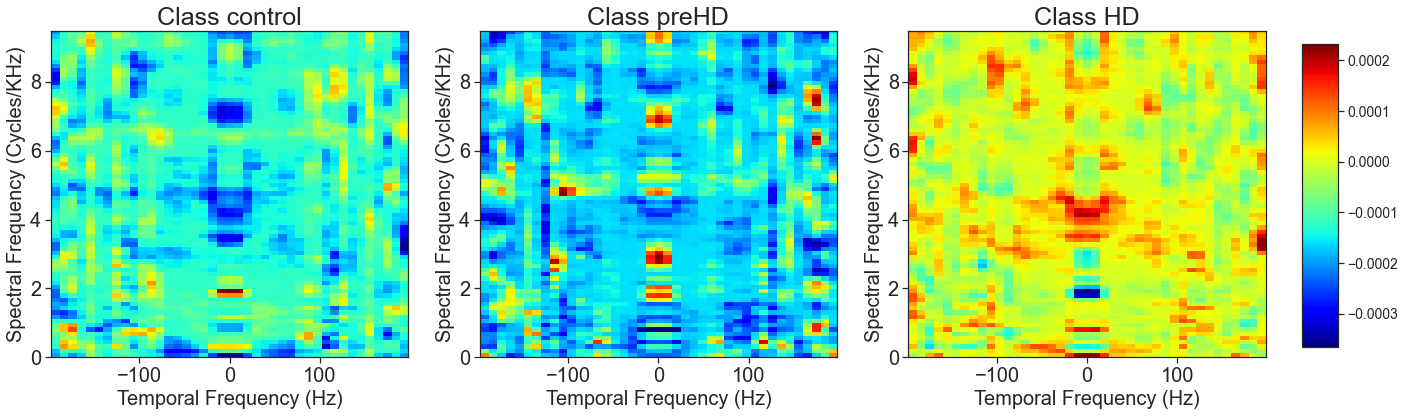}
\vspace{-0.5em}
    \label{fig:MPS}
\end{figure*}



\begin{figure}[!ht]
    \centering
    \caption{Confusion matrices for the Logistic regression based on the Phonatory features (left) and on the MPS features (right) averaged across all the repeated learning-testing experiments.}
    \includegraphics[width=0.45\textwidth]{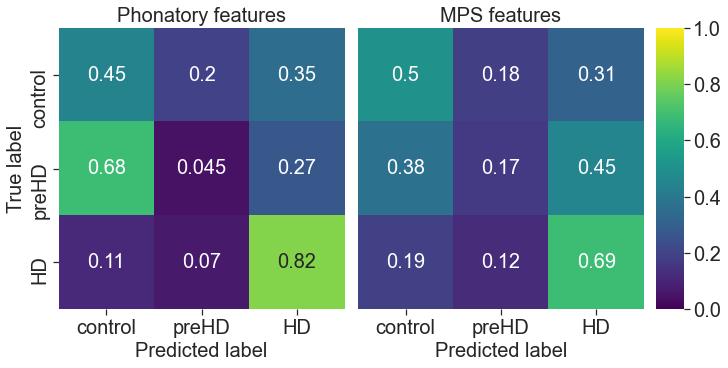}

    \label{fig:confusion}
    \vspace{-1em}
\end{figure}

\begin{table*}[!ht]
	\centering
		\caption{Results of the machine learning experiments based on the set of features obtained from the sustain of the vowel /a/. Mean and Standard Deviation are reported for each metrics. Classification performance is reported with Accuracy and F1-macro score. Regression performance is reported with the Mean Absolute Error (MAE) and the Coefficient of Determination ($R^2$). Best score for each metric is reported in \textbf{bold}.
$\dagger$ The Phonatory features do not include the tremor features as they could not be computed for each subject.
}
	\begin{tabular}{l|c|c|c|c|c|c|c|c}
		\hline
		  & \multicolumn{2}{c|}{Classification Results}   & \multicolumn{6}{c}{Regression Results}\\
		\hline

		                          & \multicolumn{2}{c|}{C vs preHD vs HD} &\multicolumn{2}{c|}{cUHDRS}    &\multicolumn{2}{c|}{TFC} &\multicolumn{2}{c}{TMS}      \\
		                          \hline

	     & Accuracy & F1-macro & MAE & $R^2$ & MAE & $R^2$ & MAE & $R^2$ \\
	    \hline
	    \hline
		Random Prior & 0.38 (0.1) & 0.31 (0.1) & 4.21 (0.6) & 0.0 (0.0) & 1.96 (0.3) & 0.0 (0.0) & 19.07 (2.3) & 0.0 (0.0)\\

		\hline
				 Phonatory feat.$\dagger$ & \textbf{0.56 (0.1)} & 0.40 (0.1) &  \textbf{2.78 (0.5)} & \textbf{0.58 (0.2)} & \textbf{1.64 (0.3)} & \textbf{0.37 (0.2)} &\textbf{13.14 (1.7)}  & \textbf{0.53 (0.2)} \\
				 \hline
				  MPS &0.54 (0.1) & 0.43 (0.1) &3.68 (0.5) & 0.28 (0.1) & 1.83 (0.3) & 0.15 (0.1) & 17.87 (3.2) & 0.26 (0.2) \\
				  \hline
				 MPS+Phonatory feat.$\dagger$ &\textbf{ 0.56 (0.1)} & \textbf{0.46 (0.1)}   & 3.10 (0.5) & 0.49 (0.1) & 1.70 (0.3) & 0.25 (0.1) & 15.34 (3.1) & 0.29 (0.2) \\
		\hline
		\hline

	\end{tabular}

	    \label{tab:results_ML}
\vspace{-1em}
\end{table*}

Statistical Analyses in Table \ref{tab:results_stats} showed that the most affected dimensions in HD are those related to airflow insufficiency and articulatory deficiency. We did not find any significant difference between PreHD and Controls, while both groups differed from HD. Post-hoc analyses revealed that HD differed from from Controls but not from preHD for pitch/F0 related features: the Degree of Pitch Breaks, the Degree of Vocal Arrests, and Standard deviation of the F0. This suggests an impairment of the vocal cords control  prior to the clinical onset. None of the Phonatory features were sufficient to distinguish preHD from Controls. Tremor features displayed the strongest effect size when comparing  preHD and Controls. But the unequal level of data loss in each group (51\% HD, 38\% preHD, and 25\% Controls) suggests defining estimation methods that incorporate tremor instability and better tremor tracking methods \cite{peplinski2019objective} for future work.

The classification results (Table \ref{tab:results_ML}) for the Phonatory features yielded lower discrimination than the MPS features. Phonatory features performances are lower than the studies \cite{rusz2014phonatory,rusz2013objective}. This can be explained by the difference in the definition of the preHD group: the preHD population were based on a previous definition (the Diagnostic Confidence Level, item 17 of the UHDRS Motor Assessment). Some of their preHD genetic carrier showed some motor deficits (TMS than was up to 8 in their population). These differences may be due to the small sample size, inter-individual variability. The differences can also be attributed to the difficulty to tease apart the 3 groups in comparison to 2 groups comparison only. Besides, the MPS features are more suitable to identify the preHD.
We also show the type of errors made by each set of features with the Confusion Matrices (see Figure \ref{fig:confusion}). HD are the most identifiable group. Clearly, the preHD are often confused by the model as HD or Controls. Yet the types of errors differ.
With the Phonatory features the preHD are even more classified as Controls than the Controls themselves (0.68 versus 0.45), which suggests a compensation mechanism.

The weights of the logistic regression trained to classify the sub-group based on the MPS features are interpretable.
These weights can be visualised in the Figure \ref{fig:MPS}. Even though the models had no prior how close the features are in the MPS space, we saw the emergence of patterns.
The HD sub-classifier showed an area of activations for spectral modulation around 4 cycles/kHZ, which can be associated with temporal modulations between -45 and 45 Hz. We found the strongest activation at the origin (0.0, 0.0), which relates to voice breaks during the phonation.
Even though the classifier is not perfect for preHD, we saw several specific activations along the Spectral modulation at ~2.5 Cycles/kHZ and ~7 Cycles/kHZ at Temporal Frequency equal to 0. This might suggest Frequency modulations specific to preHD, trying to avoid the zone around 4 Cycles/kHz of HD.
The combination of the set of MPS and Phonatory features improved the classification performances up to an Accuracy of 0.56 and F1-macro of 0.46.

In contrast, Phonatory features better reflected the clinical scores cUHDRS, TFC, and TMS (Table \ref{tab:results_ML})  The composite measure cUHDRS, currently used in the assessment of international clinical trial \cite{tabrizi2019targeting}, is the best predicted among the scores, if we rank them based on the coefficient of determination $R^2$. This means that the Phonatory features are a better indicator of the severity of the disease, once the clinical onset is declared.

\section{Conclusions}
Here, we combine the data from three groups for the study of the vocal markers of sustained phonation in Huntington's Disease patients: symptomatic, pre-symptomatic and control. We applied a statistical analysis, a classification study and assessed the capabilities to predict clinical scores. In addition, we introduced Modulation Power Spectrum features, to more traditional Phonatory features.
Airflow insufficiency and articulatory deficiency measures distinguished HD patients from both preHD and Controls.
However, Modulation Power Spectrum features provided more hope of distinguishing preHD from Controls. They allowed a three-fold reduction in misidentification of preHD.
When replicated in a larger scale and in another population, this suggest that speech phonation might replace long traditional assessments, considering that the the sustained vowel task takes less than 1 minute and UHDRS takes a minimum of 30 minutes when ran by experts. It may allow repetitive testing with limited retest effect and recordings could also be blindly scored and analysed. This points to speech as a major future tool in the clinical panel of assessments.
\section{Acknowledgements}
We  are  very thankful  to  the  patients that participated in our study. We thank Agnes Sliwinski, Katia Youssov, Laurent Cleret de Langavant for the multiple helpful discussions and the evaluations of the patients.
This work is funded in part by the Agence Nationale pour la Recherche (ANR-17-EURE-0017 Frontcog, ANR-10-IDEX-0001-02 PSL*, ANR-19-P3IA-0001 PRAIRIE 3IA Institute) and Grants from Neuratris, from Facebook AI Research (Research Gift), Google (Faculty Research Award), Microsoft Research (Azure Credits and Grant), and Amazon Web Service (AWS Research Credits).


\newpage
\begin{figure*}[!ht]
    \centering
    \caption{Results of the statistical analyses for the Modulation Power Spectrum Features between the three groups for the sustain of the vowel /a/: Controls (C), asymptomatic genetic carrier of Huntington's Disease (preHD), symptomatic genetic carrier of Huntington's Disease (HD). Left figure is the distribution of the H-statistic. Middle figure is the distribution of the uncorrected p-values. Right figure is the distribution of the FDR corrected p-values.
    }
    \includegraphics[width=0.9\textwidth]{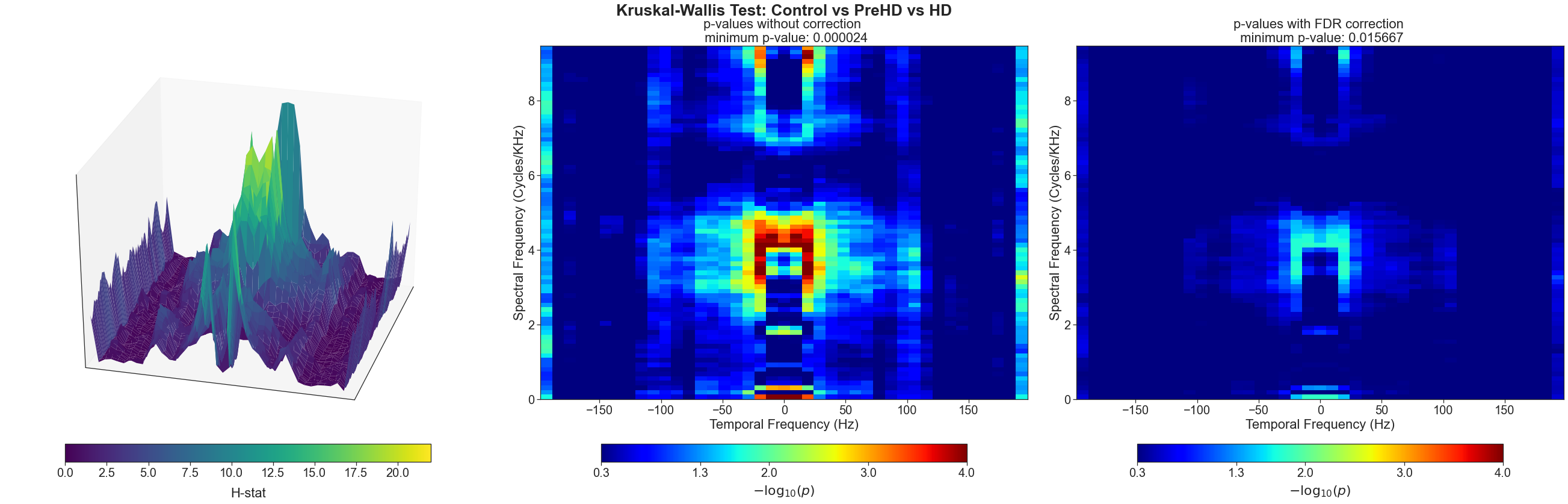}
    \label{fig:MPS_stat}
\end{figure*}

\bibliographystyle{IEEEtran}

\bibliography{mybib}

\newpage

\section{Appendix}
\subsection{Additional Statistical analysis for the Modulation Power Spectrum Features}

We tested the differences between groups with the non-parametric Kruskal-Wallis test. The False-Discovery-Rate correction is applied to correct for the multiple comparisons ($77\times 41 = 3157$). We reported the statistical results in the Figure \ref{fig:MPS_stat}.
$1.7\%$ ($14.6\%$) of the MPS features are found significant with (without) the FDR correction.
We found significant area to separate the three groups
of activation, especially around 4 cycles/kHz (very similar to the area found to identify the symptomatic HD patients).
\subsection{Interpretation of coefficients of linear models for the regression of clinical scores based on the Phonatory features}
\begin{figure}[!ht]
\vspace{-0.5em}
    \centering
    \caption{Coefficient Importance of the different Phonatory Features across the different cross-validation folds to predict the cUHDRS
    }
    \includegraphics[width=0.45\textwidth]{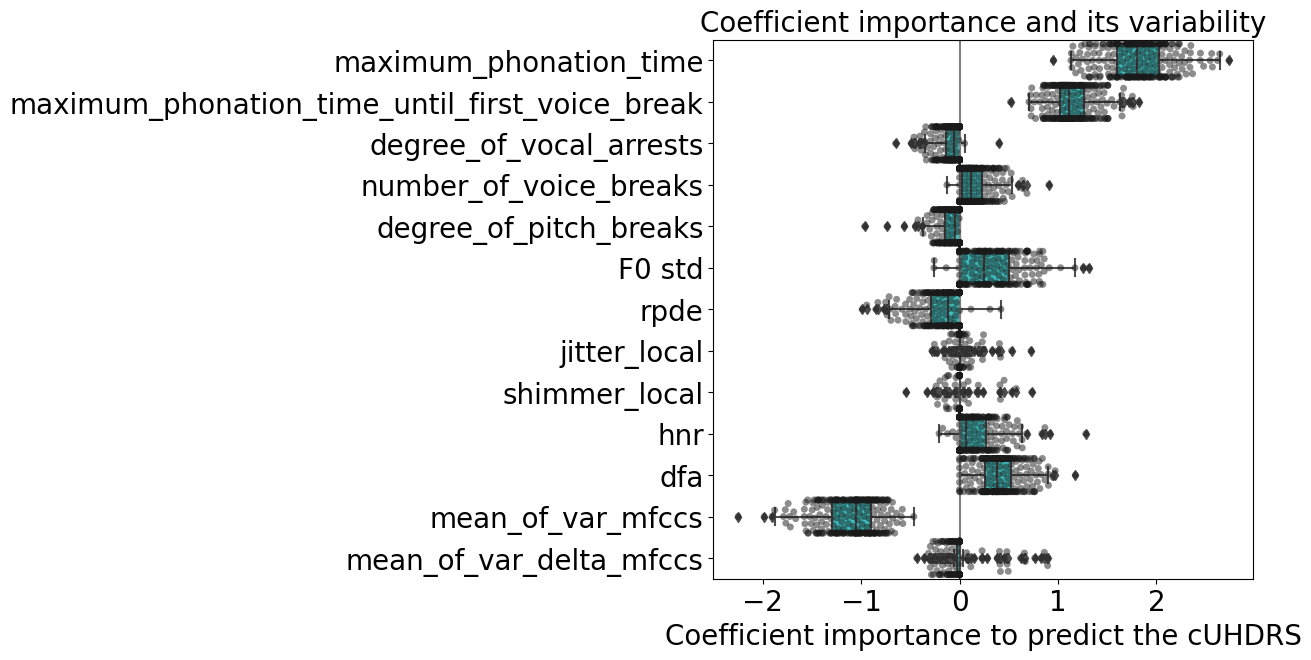}
\vspace{-0.5em}
    \label{fig:coeff_cUHDRS}
\end{figure}

\begin{figure}[!ht]
\vspace{-0.5em}
    \centering
    \caption{Coefficient Importance of the different Phonatory Features across the different cross-validation folds to predict the TFC
    }
    \includegraphics[width=0.45\textwidth]{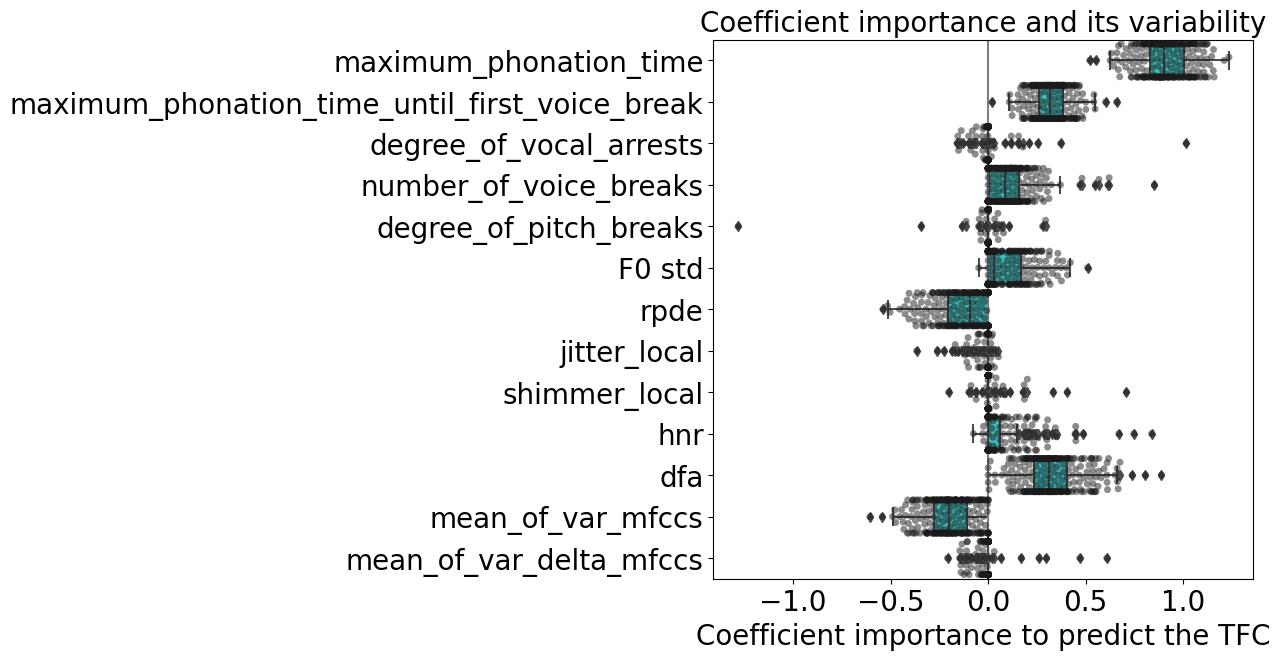}
\vspace{-0.5em}
    \label{fig:coeff_TFC}
\end{figure}

\begin{figure}[!ht]
\vspace{-0.5em}
    \centering
    \caption{Coefficient Importance of the different Phonatory Features across the different cross-validation folds to predict the TMS
    }
    \includegraphics[width=0.45\textwidth]{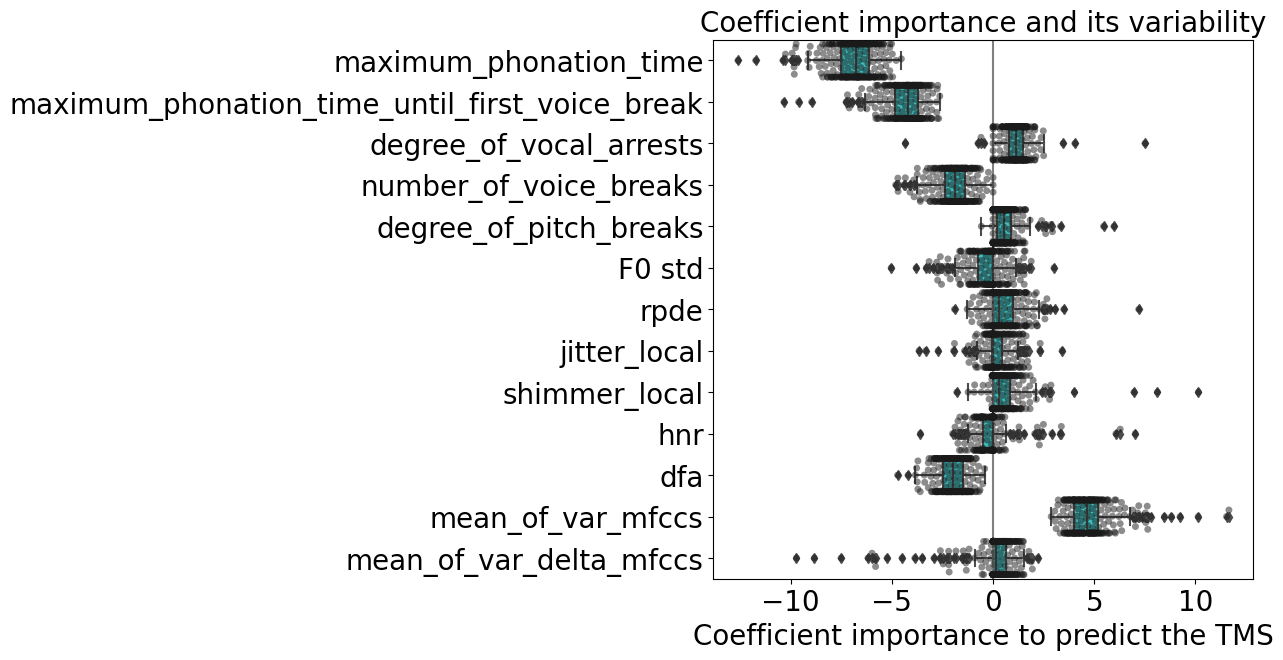}
\vspace{-0.5em}
    \label{fig:coeff_TMS}
\end{figure}

As we used linear models, each target value is modelled as a linear combination of the input features. We followed the methodology analysis from the example of scikit-learn \cite{scikit-learn} (\href{https://scikit-learn.org/dev/auto_examples/inspection/plot_linear_model_coefficient_interpretation.html}{link}). The stability of the predictors is shown through the different coefficients across folds. As we used the ElasticNet Regulariser (L1+L2 regularisation of the coefficients), we also observed the selection of variables based on the Coefficient Importance analysis. The results for the regression results for the clinical measures cUHDRS, TFC, TMS are reported respectively in the Figure \ref{fig:coeff_cUHDRS}, Figure \ref{fig:coeff_TFC}, Figure \ref{fig:coeff_TMS}.


The Maximum Phonation Time and First Occurrence of Voice Break are the features the most used for all 3 regression tasks. The Mean of SD of MFCC also contribute to the prediction of the cUHDRS and TMS. The Detrended Fluctuation Analysis is also a feature useful for the prediction of the TMS (coefficient never set to 0).

Otherwise, the contributions of the other features for the other tasks are more blurred or often set to 0. These coefficient importance give also the direction associated with the progress of the scores and then the disease.
\end{document}